\documentstyle[amssymb,twocolumn,aps,prl,epsf]{revtex}

\begin{document}

\twocolumn[\hsize\textwidth\columnwidth\hsize\csname @twocolumnfalse\endcsname   
  
\title { Singularities 
in the optical response of cuprates.}  
\author{Ar. Abanov$^1$, Andrey V. Chubukov$^1$, and  
J\"org Schmalian$^2$}   
\address{   
$^1$ Department of Physics, University of Wisconsin, Madison, WI 53706}     
\address{$^2$ Department of Physics and Ames Laboratory, 
 Iowa State University, Ames, IA 50011}   
\date{\today}   
\draft   
\maketitle    
\begin{abstract}   
 We argue that the detailed 
analysis of the optical  response in cuprate superconductors
allows one to verify the magnetic scenario of superconductivity 
in cuprates, as for strong coupling charge carriers to
 antiferromagnetic spin fluctuations, 
the second derivative of optical conductivity
 should contain detectable 
singularities at $2\Delta +\Delta_{\rm spin}$, $4\Delta$, and $2\Delta+2\Delta_{\rm spin}$, where  $\Delta$ is 
 the amplitude of the 
superconducting gap, and  $\Delta _{s}$ is the  resonance energy of spin fluctuations measured in neutron scattering. 
We argue that there is a good chance that these singularities have already been detected in the experiments on optimally doped $YBCO$. 
 \end{abstract}  
\pacs{PACS numbers:71.10.Ca,74.20.Fg,74.25.-q}   
]

\narrowtext   
The pairing state in cuprate superconductors is predominantly made out of
Cooper pairs with d$_{x^{2}-y^{2}}$-symmetry\cite{WVH93}. This salient
universal property of all
high $T_{c}$ materials entails constraints on the microscopic mechanism of
superconductivity. However, it does not uniquely determine it, leading to a
quest for experiments which can identify ''fingerprints'' of a specific
microscopic mechanism of d-wave superconductivity, a strategy similar to the
one used in conventional superconductors~\cite{a^2f}.

Several resent experiments were interpreted as an indirect evidence that d$%
_{x^{2}-y^{2}}$ pairing in cuprates is produced by an exchange of collective
spin fluctuations peaked at or near antiferromagnetic momentum ${\bf Q}=(\pi
,\pi )$~\cite{pines}. In particular, the distance between the peak and the
dip in the fermionic spectral function, $A_{{\bf k}}(\omega )$, in ARPES
experiments coincides with the frequency $\Delta _{s}$ of the resonance peak
measured in neutron scattering~\cite{ND98,ac}. This is exactly what one
should expect for fermions interacting with a resonating spin collective
mode~\cite{ND98,ac} (for phonon mediated superconductors, this is
known as the Holstein effect~\cite{holstein}). Similarly, a peak-dip
structure of the SIS tunneling conductance with peak-dip distance roughly
consistent with $\Delta _{s}$ has been obtained in the measurements on break
junctions by Zasadzinski {\em et al.} for various doping values~\cite{zasad}%
. Carbotte {\em et al.}\cite{CSB99} analyzed optical conductivity $\sigma
(\omega )$ in magnetically mediated $d$-wave superconductors and argued that 
$\Delta _{s}$ can be extracted from the measurements of the second
derivative of $\sigma (\omega )$.

In this paper we reexamine the behavior of the optical conductivity in
superconductors with quasiparticles strongly coupled to their own collective
spin modes. Our results partly agree and partly disagree with those by
Carbotte {\em et al.}\cite{CSB99} (see below). The key prediction of this
paper, however, is novel: we argue that by measuring the 
 frequency derivatives of the 
conductivity, one can not only verify the magnetic scenario, but, in
principle, also independently determine both $\Delta _{s}$ and $\Delta $ in
the same experiment.

Our argument goes as follows. For a superconductor, the real part of the 
conductivity, $\sigma_1 (\omega )$, has a 
$\delta $- functional piece due to the presence of the superconducting
condensate. A nonzero $\sigma_1 (\omega )$ at a finite frequency is only
possible if fermions have a finite lifetime. More precisely, one of the two
fermions exited in a process causing the AC conductivity should have a
finite scattering rate, while another should be able to propagate, i.e., its
energy should be larger than $\Delta $. For clean, phonon-mediated
superconductors, there are two sources for fermionic decay (see Fig. 1). 
One is a direct four-fermion interaction, which yields a threshold in the
imaginary part of the self-energy, $\Sigma ^{\prime \prime }\left( \omega
\right) $, at $\omega =3\Delta $ - the minimal energy necessary to pull all
three fermions in the final state out of the condensate of Cooper pairs.
Another is the interaction between an electron and an optical phonon. It
yields the onset of $\Sigma ^{\prime \prime }\left( \omega \right) $ at $%
\omega =\Delta +\Omega _{p}$, where $\Omega _{p} $ is the frequency of an
optical phonon~\cite{holstein} (for simplicity, we assumed that the phonon
propagator has a single pole). For the values of the coupling constant used
to interpret the tunneling data in strongly coupled conventional
superconductors like $Pb$~\cite{recentPRL}, $\Omega _{p}>2\Delta $, i.e.,
the onset of conductivity is at $3\Delta +\Delta =4\Delta $ ($2\Delta $ for
dirty superconductors~\cite{lv}), while the signatures of phonon-assisted
damping only show up at higher frequencies, and are uncorrelated with the
behavior of $\sigma_1 (\omega )$ near $4\Delta $. %
\begin{figure}[tbp]
\begin{center}
\epsfxsize=3.2in \epsfysize=0.6in
\epsffile{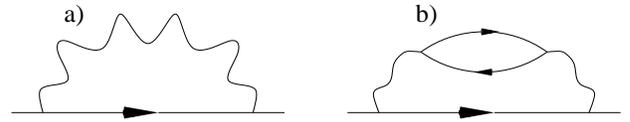}
\end{center}
\caption{a) The exchange diagram for boson mediated interaction. The solid
line stands for a propagating fermion. The wiggled line is a phonon
propagator in case of electron- phonon interaction, and a magnon line in
case of spin- fluctuation mediated interaction. b) The lowest order diagram
for the fermionic self energy due to a direct four fermion interaction, also
represented by a wiggly line}
\label{Figure1}
\end{figure}

For spin-mediated superconductivity, the situation is different. In the
one-band model for cuprates, which we adopt, the underlying interaction is
solely a Hubbard-type four-fermion interaction. Spin excitations appear as
collective modes of fermions, and their velocity $v_s$ is comparable to $v_F$%
. For $v_s \sim v_F$, the low frequency spin dynamics is dominated by a
decay process into a particle-hole pair and is purely relaxational in the
normal state, with nearly featureless $\chi ^{\prime \prime }({\bf Q},\omega
)$~\cite{ac}. (for phonon superconductors the relaxation is also present
but is strongly reduced due to a smallness of the sound velocity compared to 
$v_{F}$~\cite{mahan}).

Below $T_{c}$, fermions acquire a gap, and spin-decay becomes impossible for 
$\ $energies below $2\Delta $. The direct four-fermion interaction 
process in Fig.~\ref{Figure1}b 
then yields a threshold in $\Sigma ^{\prime \prime }$ at $3\Delta $ which
gives rise to a singularity in the conductivity at $\omega =4\Delta $~\cite{scalapino}. If $%
\chi^{\prime\prime} ({\bf Q},\omega )$  remained featureless, this would be
the only effect. However, several authors have demonstrated that the
residual attraction in a $d$-wave superconductor 
binds a particle and a hole into a spin exciton at an energy $\Delta
_{s}<2\Delta $. This effect gives rise to a peak in $\chi^{\prime\prime} (%
{\bf Q},\omega )$ at $\omega =\Delta _{s}$ and makes it look like the
spectral function for optical phonons. Accordingly, the conductivity
acquires another threshold at $2\Delta +\Delta _{s}$. Formally, this is
analogous to the phonon case, but in distinction to phonons, $\Delta
_{s}<2\Delta $. Then $2\Delta +\Delta _{s}<4\Delta $, i.e., in clean
systems, the lower threshold corresponds to the scattering by a spin
exciton. Moreover, since both effects are due to the same underlying
interaction, (the  diagram in Fig.~\ref{Figure1}b is just
the first term in the series of graphs which constitute the spin-mediated
scattering process shown in Fig~\ref{Figure1}b)
the ratio $\Delta _{s}/\Delta $ and the relative intensity of the
singularities in $\sigma_1 (\omega )$ at $4\Delta $ and $2\Delta +\Delta _{s}$
are correlated. This correlation is a ``fingerprint'' of the
spin-fluctuation mechanism. We will argue that there are strong indications
that both singularities have been observed in the measurements of the
optical conductivity in ${\rm YBCO}$~ \cite{CSB99}, and their position and
relative intensity are in reasonable 
agreement with the theory.

Before we proceed with the calculations, a comment is in order. In the above
discussion we neglected the momentum dependence of the $d$-wave gap.
Meanwhile, the computations of the optical conductivity involve averaging of
the lifetime over the Fermi surface~\cite{ps,IM98}. It is then {\it a'priori}
unclear whether the angular dependence of the $d$-wave gap with $\Delta
(\theta )\propto \cos 2\theta $ affects the positions of the two thresholds
in the conductivity. Carbotte {\it et al} argued ~\cite{CSB99} that it does,
and the singularity at $2\Delta +\Delta _{s}$ 
(which they only considered)  
is determined by some averaged $|2\Delta (\theta )|\approx \Delta $. We 
argue that averaging reduces  strengths of the singularities but doesn't
shift their positions. Our argument is two-fold. First, we explicitly
demonstrate below that the singularity in the conductivity occurs at a
frequency equal to the maximum value of the gap. Second, two of us and
Finkel'stein argued earlier~\cite{acf} that for spin-mediated $d$-wave
superconductivity, $\Delta (\theta )$ is at its maximum at hot spots (points
at the Fermi surface separated by ${\bf Q}$)~\cite{acf}. These are precisely
the Fermi points 
which determine the position of the excitonic pole in $\chi ^{\prime \prime
}({\bf Q},\omega )$. Accordingly, the singularity in conductivity entirely
comes from fermions near hot spots, and the threshold frequency $2\Delta
+\Delta _{s}$ involves a maximum value of the gap {\it and} the resonance
spin frequency at momentum ${\bf Q}$.  The same argumentation implies that $%
4\Delta $ threshold also involves a maximum value of the gap. Note for
clarification that we are only considering here the singularities in the
conductivity at $\omega >2\Delta $. The regular part of
 $\sigma_1 (\omega)$ is not necessary confined to hot spots. In particular,
 for $\omega \ll \Delta $ the optical
response is dominated by nodal quasiparticles for which $\Sigma ^{\prime
\prime }$ is nonzero down to the lowest frequencies.

We now proceed with the calculations. The real part of the 
optical conductivity in a
superconductor is given by 
\begin{equation}
\sigma_1 (\omega )={\rm Re}\frac{i}{\omega +i\delta }~\int d\theta \Pi
_{\sigma }(\theta ,\omega )  \label{rs}
\end{equation}
where $\Pi _{\sigma }(\theta ,\omega )$ is the fully renormalized
current-current correlator. In Matsubara frequencies, it is given by
\begin{eqnarray}
\Pi _{\sigma }\left( i\omega _{n}\right)  &\propto &\frac{1}{\beta }%
\sum_{m}\int \frac{d^{2}k}{\left( 2\pi \right) ^{2}}\left[ G_{{\bf k}}\left(
i\omega _{n}+i\omega _{m}\right) \text{ }G_{{\bf k}}\left( i\omega
_{m}\right) \right.   \nonumber \\
&&\left. +F_{{\bf k}}\left( i\omega _{n}+i\omega _{m}\right) \text{ }F_{{\bf %
k}}\left( i\omega _{m}\right) \right] ,  \label{bub}
\end{eqnarray}
and the normal and anomalous Green's functions are  
\begin{eqnarray}
G_{{\bf k}}(i\omega _{m}) &=&\frac{\Sigma _{{\bf k}}(i\omega
_{m})+\varepsilon _{{\bf k}}}{\Sigma _{{\bf k}}^{2}(i\omega _{m})-\Phi _{%
{\bf k}}^{2}(i\omega _{m})-\varepsilon _{{\bf k}}^{2}}, \\
F_{{\bf k}}(i\omega _{m}) &=&\frac{\Phi _{{\bf k}}(\omega _{m})}{\Sigma _{%
{\bf k}}^{2}(i\omega _{m})-\Phi _{{\bf k}}^{2}(i\omega _{m})-\varepsilon _{%
{\bf k}}^{2}},  \label{GF}
\end{eqnarray}
(we adsorbed a bare $i\omega _{m}$ term into $\Sigma _{{\bf k}}(i\omega _{m})
$. In principle, $\Pi _{\sigma }$ is modified by vertex corrections related
to  $d\Sigma /dk$~\cite{mahan}, but for spin-mediated scattering these corrections are small (see below). 

As an input for the computation of $\Pi _{\sigma }$  
we need the forms of the fermionic
self-energy $\Sigma _{{\bf k}}(i\omega _{m})$ and the anomalous vertex $\Phi
_{{\bf k}}(i\omega _{m})$. We obtained these forms in Ref.\cite{ACS00} by
deriving and solving a set of Eliashberg equations within the spin-fermion
model. This model adequately describes the interaction between low-energy
fermions and their collective spin degrees of freedom\cite{ac,acf,ACS00} at
energies smaller than $E_{F}$. The full dynamical spin susceptibility peaked
at (or near) ${\bf Q}$ mediates $d_{x^{2}-y^{2}}$ superconductivity. As
discussed, this susceptibility is by itself affected by low-energy fermions
via a decay process into a particle and a hole,  and has to be computed
together with the fermionic self-energy and the pairing vertex.

The justification of the Eliashberg approach for the spin-mediated
superconductivity was discussed earlier~\cite{ac,acf,ACS00}, and we just
quote the result: at strong dimensionless spin-fermion coupling $\lambda$,
vertex corrections and $v^{-1}_F~d \Sigma/d k_\perp$, where $k_\perp$ is the
component of the momentum transverse to the Fermi surface, are small
compared to $d\Sigma/d\omega$ by $\log \lambda/\lambda$. In what follows we
will neglect these corrections, i.e., approximate $\Sigma _{{\bf k}%
}(i\omega_m)$ and $\Phi_{{\bf k}}(i\omega_m)$ by $\Sigma _{{\bf k}%
}(i\omega_m) = \Sigma (i\omega_m, \theta)$ and $\Phi (i\omega_m,\theta)$.

\begin{figure}[tbp]
\centerline{\epsfxsize=\columnwidth 
\epsfysize=0.6\columnwidth 
\epsffile{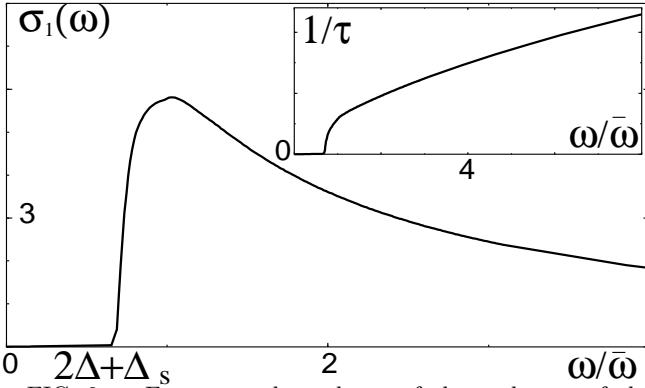}}
\caption{ Frequency dependence of the real part of the 
optical conductivity $\sigma_1 (\omega)$ at $T=0$ computed using
the self energy and the pairing vertex determined from the Eliashberg
equations for $\lambda =1$. The onset of the optical response is $%
\protect\omega =2\Delta +\Delta _{s}$. The contributions from 
nodal regions
 (not included in
calculations) yield a nonzero conductivity at all $\protect\omega$. 
They
also soften the singularity at $\protect\omega =2\Delta +\Delta _{s}$, but
do not eliminate it.
Insert - the behavior of the relaxation rate $1/\tau (\omega) = (4\pi/\omega^2_{pl})
 Re[1/\sigma (\omega)]$. The frequency is measured in units ${\bar \omega}$ which sets
the overall energy scale in the Eliashberg solution. For $\lambda =1$, 
$\Delta = 0.204 {\bar \omega}$ and $\Delta_s = 0.291 {\bar \omega}$.}
\label{Figure_cond_strongc-th}
\end{figure}

As our goal is to study the singularities in $\sigma_1 (\omega )$, we first
perform calculations assuming that $\Sigma $ and $\Phi $ are independent on $%
\theta $ (i.e. that the superconducting gap is flat near the hot spots), and
then analyze the results for a true $d$-wave gap. For a flat gap, the
momentum integration in Eq. (\ref{bub}) is straightforward. Substituting $k$
integration by integration over $\varepsilon _{k}$, and performing it, we
obtain at $T=0$ and $\omega \neq 0$ 
\begin{equation}
\Pi _{\sigma }\left( i\omega _{n}\right) \propto \int d\omega _{m}^{\prime
}d\theta \frac{\Sigma _{+}\Sigma _{-}+\Phi _{+}\Phi _{-}+D_{+}D_{-}}{%
D_{+}D_{-}(D_{+}+D_{-})}  \label{sigma}
\end{equation}
Here $\Sigma _{\pm }=\Sigma \left( i\omega _{\pm },\theta \right) $, $\Phi
_{\pm }=\Phi \left( i\omega _{\pm },\theta \right) $, and $D_{\pm }=(\Phi
_{\pm }^{2}-\Sigma _{\pm }^{2})^{1/2}$, where $\omega _{\pm }=\omega
^{\prime }\pm \omega /2$. The conductivity is obtained by converting this
expression to the real axis~\cite{dolgov}.
The singular piece in $\sigma_1 (\omega )$ near $2\Delta +\Delta _{s}$ can be
obtained without a precise knowledge of $\Sigma (\omega )$ and $\Phi (\omega
)$: the only information we need is that in a $d$-wave superconductor, $\chi
^{\prime \prime }({\bf Q},\omega )$ has a $\delta $-functional singularity
at $\omega =\Delta _{s}$. This is what we found solving a set of three
Eliashberg equations. Using this as an input and applying a spectral
representation for $\Sigma ^{\prime \prime }$ and $\Phi ^{\prime \prime }$ 
(which for a given $\chi^{\prime \prime} (Q,\omega)$ are described by a conventional set of two 
Eliashberg equations), 
we obtain that $\Sigma ^{\prime \prime }(\omega )$ and $\Phi ^{\prime \prime
}(\omega )$ are zero up to $\omega =\Delta +\Delta _{s}$, and undergo finite
jumps at this frequency. By Kramers-Kronig relation, $\Sigma ^{\prime }$ and 
$\Phi ^{\prime }$ diverge as $|\log (\omega -\omega _{0})|$ where $\omega
_{0}=\Delta +\Delta _{s}$. The prefactor is the same for $\Sigma ^{\prime }$
and $\Phi ^{\prime }$. Substituting these forms of $\Sigma (\omega )$ and $%
\Phi (\omega )$ into (\ref{bub}) we obtain after simple algebra that the
conductivity emerges above $2\Delta +\Delta _{s}$ as $\epsilon ^{1/2}/\log
^{2}\epsilon $, where $\epsilon =\omega -(2\Delta +\Delta _{s})$. This
singularity obviously causes a divergence in the derivatives of the
conductivity at $\epsilon =+0$.

In Fig.\ref{Figure_cond_strongc-th} we show the result for the conductivity
obtained by numerically solving Eq.(\ref{bub}) using $\Sigma (\omega )$ and $%
\Phi (\omega )$ from~Ref.\cite{ACS00}. We clearly see the expected threshold
at $2\Delta +\Delta _{s}$. The insert shows the behavior of the
 relaxation rate $1/\tau (\omega) =  (4\pi/\omega^2_{pl})
 Re[1/\sigma (\omega)]$ where $\omega_{pl}$ is the plasma frequency.
Observe that $1/\tau (\omega)$  is linear in 
$\omega $ over a rather wide frequency range. This agrees with the earlier 
study of the normal state conductivity~\cite{rob}.

\begin{figure}[tbp]
\centerline{\epsfxsize=\columnwidth 
\epsfysize=0.6\columnwidth 
\epsffile{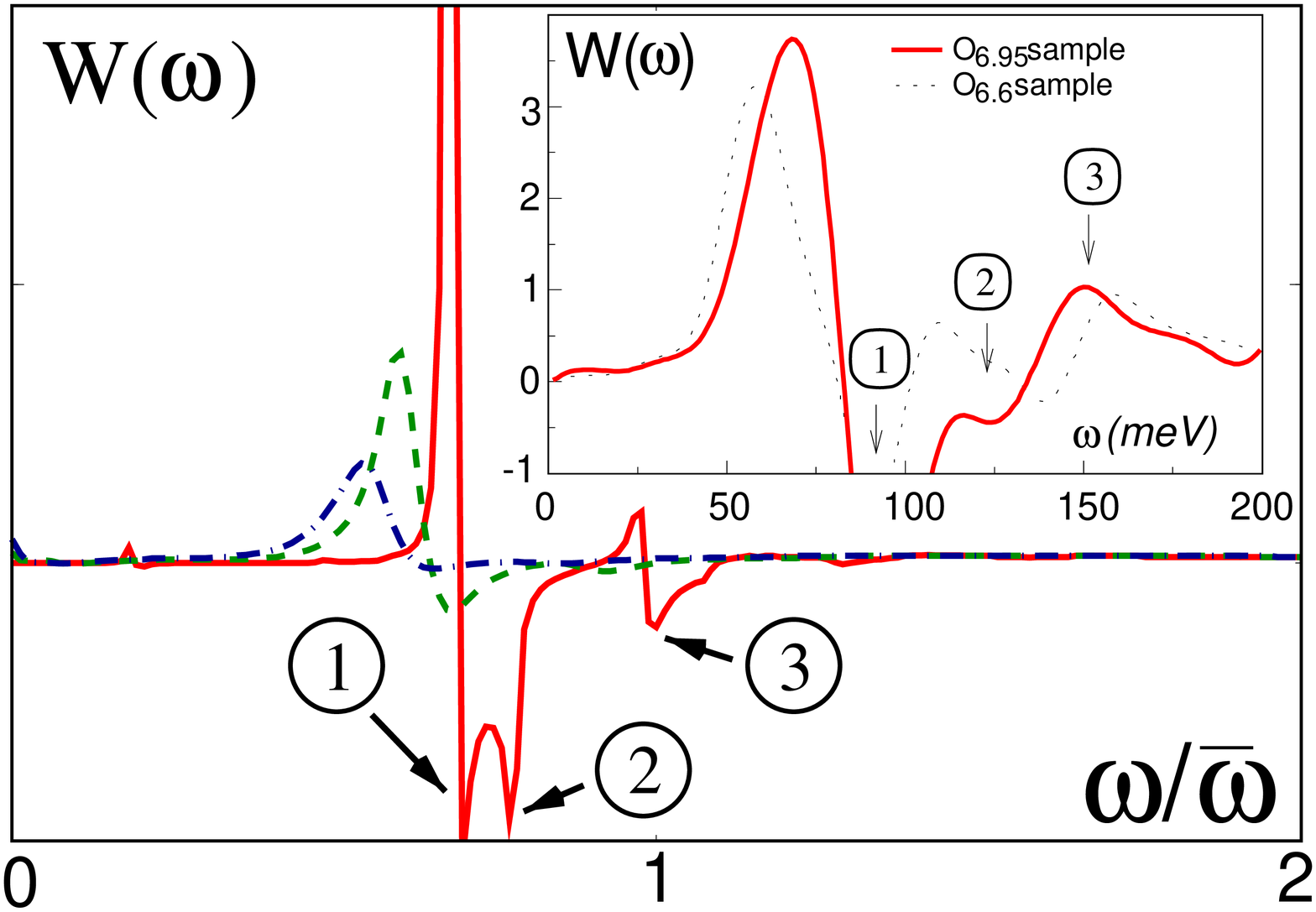}}
\caption{  A calculated frequency dependence of $W(\protect\omega )=\frac{%
d^{2}}{d^{2}\protect\omega }[\protect\omega {\rm Re}[\protect1/\sigma(%
\protect\omega )]]$ at $T \rightarrow 0$. This quantity is a sensitive measure of fine structures
in the optical response. The locations of the extrema are: 1--$2\Delta + \Delta_s$, 2--$4\Delta$, 3--$2\Delta + 2\Delta_s$.
Dashed lines are the results are higher $T$. Observe that
the maximum shifts to a lower temperature, but the 
 minimum remains at $2\Delta +\Delta _{s}$.
Inset-- Experimental results for $W(\protect\omega )$ at low $T$ 
 from Ref.\protect\cite{CSB99}. The position of the deep minimum agrees
well with $2\Delta +\Delta _{s}$. The extrema at higher frequencies are
consistent with $4\Delta $ and $2\left( \Delta +\Delta _{s}\right) $
predicted by the theory.}
\label{Figure_d2sigdw2}
\end{figure}

We next demonstrate that the position of the singularity is not affected by
the angular dependence of the gap. Indeed, let the maximum value of the gap
correspond to $\theta =0$ and symmetry related points. At deviations from $%
\theta =0$ both $\Delta $ and $\Delta _{s}$ decrease. The decrease of $%
\Delta $ is obvious, the decrease of $\Delta _{s}$ is due to the fact that
resonance is a feedback from superconductivity, and its frequency scales as $%
(\Delta (\theta))^{1/2}$. Since both $\Delta $ and $\Delta _{s}$ are maximal
at a hot spot, we can expand $\omega _{0}(\theta )=\Delta (\theta )+\Delta
_{s}(\theta )$ as $\omega _{0}(\theta )=\omega _{0}-a{\theta}^{2}$, where $%
a>0$. The singular pieces in $\Sigma (\omega)$ and $\Phi (\omega)$  then
behave as $|\log (\omega _{0}-\omega -a\theta ^{2})|$. Substituting these
forms into (\ref{bub}) and integrating over $\theta $, we find that the
conductivity itself and its first derivative are continuous at $\omega
=2\Delta +\Delta _{s}$, but the second derivative of the conductivity
diverges as ${d^{2}\sigma }/{d\omega ^{2}}\propto 1/(|\epsilon |\log
^{2}\epsilon )$ where, we remind, $\epsilon =\omega -(2\Delta +\Delta _{s})$%
. We see that the singularity is weakened by the angular dependence of the
gap, but it is still located at exactly $2\Delta +\Delta _{s}$.

The same reasoning is also applied to a region near $4\Delta$. We found that
the singularity at $4\Delta$ is also weakened by the angular dependence of
the gap, but is not shifted and still should show up in the second
derivative of the conductivity.

We now discuss the second derivative of the conductivity in more detail. 
In Fig.\ref{Figure_d2sigdw2} we present our numerical results for $W(\omega
)=\frac{d^{2}}{d^{2}\omega }(\omega {\rm Re}\sigma ^{-1}(\omega ))$ 
 which effectively measures second derivative of conductivity
(we followed Ref~\cite{CSB99} and used the same $W(\omega )$ as for phonon
superconductors). We clearly see that there is a sharp maximum in $W(\omega )
$ near $2\Delta +\Delta _{s}$ followed by a deep minimum. We also see that $%
W(\omega )$ has extra extrema at $4\Delta $ and, also, at $2\omega
_{0}=2\Delta +2\Delta _{s}$. The last peak is a secondary effect due to a
singularity in $\Sigma (\omega )$ at $\omega =\omega _{0}$: $\sigma_1 (\omega )
$ is singular when the frequencies of {\it both} fermions in the
polarization bubble exceed $\omega _{0}$.

The experimental result for $W(\omega )$ in YBCO is shown in the insert.
 We see that the theoretical and experimental plots of $%
W(\omega )$ look rather similar, and the relative intensities of the peaks
are at least qualitatively consistent with the theory. By the reasons which
we display below, we identify $2\Delta +\Delta _{s}$ with the deep minimum
in $W(\omega )$. This yields $2\Delta +\Delta _{s}\approx 100{\rm meV}$.
Identifying the extra extrema in the experimental $W(\omega )$ with $4\Delta 
$ and $2\Delta +2\Delta _{s}$, respectively, we obtain $4\Delta \sim 130{\rm %
meV}$, and $2\Delta +2\Delta _{s}\sim 150{\rm meV}$. We see that three sets
of data are self-consistent and yield $\Delta \sim 30{\rm meV}$ and $\Delta
_{s}\sim 40-45{\rm meV}$. The value of $\Delta $ is in good agreement with
tunneling measurements~\cite{Miyakawa99}, and $\Delta _{s}$ agrees well with
the resonance frequency extracted from neutron measurements~\cite{neutrons}.
We caution, however, that determination of a second derivative of a measured
quantity is a very subtle procedure. The good agreement between our theory
and the experiment is promising but have to be verified in further
experimental studies.
Nevertheless, our calculation clearly demonstrates the presence and
observability of these ''higher harmonics'' of the optical response at $%
4\Delta $ and $2\Delta +2\Delta _{s}$.

\begin{figure}[tbp]
\centerline{\epsfxsize=\columnwidth 
\epsfysize=0.6\columnwidth 
\epsffile{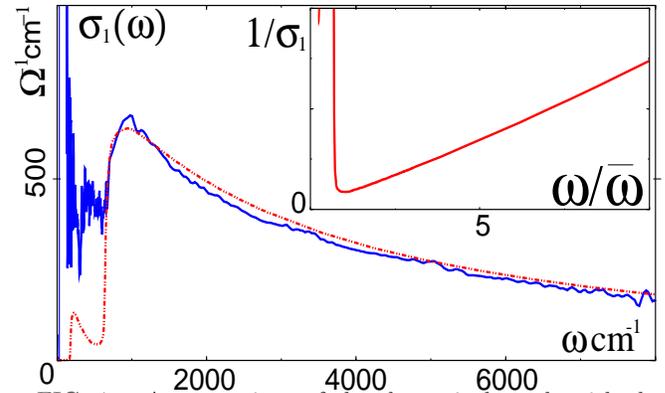}}
\caption{
A comparison of the theoretical result with the experimental
data of Puchkov {\it et al.}~\protect\cite{DNB96}. The substructure
in the theoretical $\sigma_1$ at low frequencies is an artifact.
Insert - the behavior of $\protect\sigma_1^{-1} (\protect%
\omega)$.}
\label{Figure_cond_strongc-exp}
\end{figure}

So far we considered only the singular part of $\sigma_1 (\omega )$. In Fig.%
\ref{Figure_cond_strongc-exp} we compare our results for $\sigma_1 (\omega )$
(ignoring the contributions from the nodes) directly with the experimental
data by Puchkov {\em et al.}\cite{DNB96} for optimally doped YBa$_{2}$Cu$_{3}
$O$_{6+\delta }$. We used 
$\lambda =1$ and
 the overall energy scale ${\bar \omega}$
 which yield $\Delta \sim 30{\rm meV}
$ and $\Delta _{s}\sim 45{\rm meV}$ as the solution of the Eliashberg set,
and also $\omega_p = 1.2\times 10^{4} cm^{-1}$, similar to that in~\cite{DNB96}.
We see that the frequency 
dependence of the conductivity at high frequencies agrees well 
with the data. The
measured conductivity drops at about $100{\rm meV}$ in rough agreement with $%
2\Delta +\Delta _{s}\approx 100{\rm meV}$ in our theory. 
As in earlier studies~~\cite{IM98,rob}, to match the
magnitude of the conductivity, we had to add the constant 
$7 \times10^{-4} \Omega$ 
{\rm cm} to $(\sigma_1 (\omega ))^{-1}$.
 We view 
the good agreement between theory
and experiment at $\omega > 2\Delta + \Delta_s$ is predominantly an indication  that the momentum dependence of the fermionic dynamics becomes
irrelevant at high frequencies, and fermions from all over the Fermi surface
behave as if they were at hot spots. The insert to 
Fig.\ref{Figure_cond_strongc-exp} shows $\sigma_1^{-1} (\omega)$. 
We see that it is linear over a substantial frequency range. 

We emphasize, however,
 that this linearity is only an intermediate asymptotic. At
the highest $\omega $, our theory  
yields $\sigma_1 (\omega )\propto \omega ^{-1/2}$. 
The lower boundary for $\omega^{-1/2}$ behavior 
decreases with increasing $\lambda$. 
For optimally doped $Bi2212$, $\lambda$ is somewhat larger than in $YBCO$ 
as $\Delta \approx \Delta_s$, and we expect a more pronounced $\omega^{-1/2}$ behavior at high frequencies.  This trend is
 consistent with the  data of Ref.\cite{Brookhaven}. This
 issue, however, requires further study as 
 $\sigma_1 (\omega )\varpropto \omega ^{-1/2}$ at intermediate frequencies 
was also obtained in Ref.\cite{IM98} 
assuming a strongly momentum dependent scattering rate. 

Finally, we comment on the position of the $2\Delta +\Delta _{s}$ peak and
compare our results with those by Carbotte {\em et al.}\cite{CSB99}.
Theoretically, at $T=0$ and in clean limit, the maximum and minimum in $%
W(\omega )$ are at the same frequency. We found, however, that at finite $T$%
, they quickly move apart. We present the theoretical temperature dependence of $%
W(\omega )$ in Fig~\ref{Figure_d2sigdw2}.
Carbotte {\em et al.}~\cite{CSB99} focused on the maximum in $W(\omega )$
and argued that it is located at $\Delta +\Delta _{s}$ instead of $2\Delta
+\Delta _{s}$. We also found that the maximum in $W(\omega )$ shifts to a
lower frequency with increasing temperature, already at $T$ where the
temperature dependence of the gap may be neglected. On the other hand, the
minimum in $W(\omega )$ moves very little with increasing $T$ and virtually
remains at the same frequency as at $T=0$. This is our reasoning to use the
minimum in $W(\omega )$ as a much more reliable feature for the comparison
with experiments. 
This reasoning is in agreement with recent conductivity data on 
optimally doped $Bi2212$~\cite{Brookhaven} -- $W(\omega )$ extracted from these data shows
strong downturn variation of the maximum in $W(\omega )$ with increasing
temperature, but the minimum in $W(\omega )$ is located at around $110meV$
for all temperatures.

Finally, we briefly consider whether one can extract a resonance spin frequency from the measurements of the Raman intensity in a $d-$wave superconductor. 
The Raman intensity  
is given by~\cite{klein} 
\begin{equation}
R (\omega) = {\rm Im} \int d \theta V^2 (\theta) \Pi_R (\omega, \theta)
\end{equation}
where $V(\theta)$ is Raman matrix element, and 
$ \Pi_R$ is same bubble as for conductivity, but with 
a different sign of the anomalous $FF$ term. The latter is a consequence
of the fact that Raman vertices are scalar and do not change sign under $k -> -k$. Performing the integration over quasiparticle energies in the same way as for conductivity we obtain in Matsubara frequencies~\cite{girsh}
 \begin{equation}
\Pi _{\sigma }\left( i\omega _{n}\right) \propto \int d\omega _{m}^{\prime
}d\theta \frac{\Sigma _{+}\Sigma _{-}-\Phi _{+}\Phi _{-}+D_{+}D_{-}}{%
D_{+}D_{-}(D_{+}+D_{-})}  \label{r1}
\end{equation}
 For
mostly studied $B_{1g}$ scattering, the Raman vertex has the same angular
dependence as the $d$-wave gap, i.e., $V(\theta )\propto \cos \left( 2\theta
\right)$~\cite{klein,Devereaux}. 
Straightforward computations then show that
for a $d-$wave gas, 
 $R(\omega )\propto \omega ^{3}$ at low frequencies\cite{Devereaux}, and diverges as $\omega$ approaches $2\Delta$ first as $|\omega - 2\Delta|^{1/2}$, and
then, in the immediate vicinity of $2\Delta$, as $\log |\omega - 2\Delta|$~\cite{Devereaux,natan}. 
 At larger frequencies $R(\omega )$
gradually decreases.
At strong coupling, we performed the same analysis as for conductivity and
found that the sign change of the $FF$ term in the bubble, compared to that 
for conductivity, has a drastic effect:  near $2\Delta + \Delta_s$,
 singular contributions  from $\Sigma _{+}\Sigma
_{-} $ and $\Phi _{+}\Phi _{-}$ terms in  Eq.(\ref{r1}) cancel
each other. As a result,  for a flat gap, only the second
derivative of $R(\omega )$ diverges at $2\Delta +\Delta_s$. For a
quadratic variation of a gap near its maximum, the singularity is even
weaker and shows up only in the third derivative of $R(\omega )$. Obviously,
this is a much weaker effect than that 
for conductivity, and its determination requires a high quality of
the experiment. Notice, however, that due to the
closeness of hot spots to $(0,\pi )$ and related points, at which $v_{F}$
vanishes, the actual smearing of the singularity due to the 
angular integration may be
less drastic than in our theory and the singularity in $R(\omega)$ 
at $2\Delta +\Delta_s$ may
possibly be extracted from the data.

To conclude, in this paper we examined the singularities in the optical
conductivity in a $d$-wave superconductors assuming that the pairing is
mediated by overdamped spin fluctuations. We argued that $\sigma_1 (\omega )$
should have singularities at $2\Delta +\Delta _{s}$, $4\Delta $ and $2\Delta
+2\Delta _{s}$, where $\Delta $ is the maximum value of the $d-$wave gap, and 
$\Delta _{s}<2\Delta $ is the resonance spin frequency. The experimental
detection of these singularities would be a strong argument in favor of the
magnetic scenario. We argued that there
is a good possibility that all three singularities have actually been
detected in recent data on $YBCO$. 

It is our pleasure to thank D. N. Basov, G. Blumberg, J.C. Campuzano, 
J. Carbotte, P.
Coleman, O. Dolgov, P. Johnson, M. Norman, 
D. Pines,  E. Schachinger, S. Shulga and J. Zasadzinski for useful
conversations. We are also thankful to D. N. Basov, 
C. Homes, M. Strongin and J. Tu for sharing
unpublished results with us. The research was supported by NSF DMR-9979749
(Ar. A and A. Ch.) and by
 the Ames Laboratory, operated for the 
U.S. DoE by Iowa State University under contract No.
W-7405-Eng-82 (J.S).

\end{document}